\documentclass[aps,pra,showpacs,twocolumn]{revtex4}
\usepackage{amssymb}
\usepackage{amsmath}
\usepackage{graphicx}
\usepackage{epsfig}

\begin{document}

\title{Beam Splitter for Spin Waves in Quantum Spin Network}
\author{S. Yang$^{1}$, Z. Song$^{1,a}$ and C.P. Sun$^{1,2,a,b}$ }
\affiliation{$^{1}$ Department of Physics, Nankai University, Tianjin 300071, China}
\affiliation{$^{2}$ Institute of Theoretical Physics, Chinese Academy of Sciences,
Beijing, 100080, China}

\begin{abstract}
We theoretically design and analytically study a controllable beam splitter
for the spin wave propagating in a star-shaped (e.g., a $Y$-shaped beam)
spin network. Such a solid state beam splitter can display quantum
interference and quantum entanglement by the well-aimed controls of
interaction on nodes. It will enable an elementary interferometric device
for scalable quantum information processing based on the solid system.
\end{abstract}

\maketitle

\section{Introduction}

Beam splitters are the elementary optical devices frequently used in
classical and quantum optics \cite{QOP}, which can even work well in the
level of single photon quanta \cite{S-photon} and are applied to generate
quantum entanglement \cite{Entng}. For matter waves, an early beam splitter
can be referred to the experiments of neutron interference based on a
perfect crystal interferometer with wavefront and amplitude division \cite%
{N-interfer}; and now an atomic beam splitter has been experimentally
implemented on the atom chip \cite{C-atom}. The theoretical protocols have
been suggested to realize the beam splitter for the Bose-Einstein condensate
\cite{BEC}.

In this paper, we propose and study the implementation of beam splitter for
the spin wave propagations in the star-shaped spin networks (SSSNs) with $%
m+1 $ weighted legs (see the Fig. 1a), where each leg is a one-dimensional
(1-D) spin chain with $XY$ couplings. This investigation is mostly motivated
by our recent researches on the perfect transfer of quantum states along a
single quantum spin chain \cite{our-spin} and for a 1-D Bloch electron
system \cite{our-bloch,YS1}. The similar quantum networks have been
considered for a coupled many harmonic oscillator system \cite{Plenio} and
for quantum cloning via spin networks \cite{du}.

A basic SSSN is a $Y$-shaped network or called $Y$-beam \cite{C-atom} for $%
m=2$, which can be regarded as an elementary block, in principle, to the
architecture of complicated networks (such as a solid state interferometer)
for quantum information processing. It can function to transfer quantum
state coherently in multi-channel and to generate entanglement between two
spins which are a long distance apart. Furthermore, we will show that the
quantum coherence of spin waves propagating in two legs can be well
controlled by adjusting the coupling strengths only at the node; and then a
controllable solid state interferometer is built based on this crucial
function.
%%%%%%%%%%%%%%%%%%%%%%%%%%%%%%%%%%%%%%%%%%%%%%%%%%%%%%%%%%%%%%%%%%%%
\begin{figure}[tbp]
\includegraphics[bb=90 285 490 740, width=4 cm,clip]{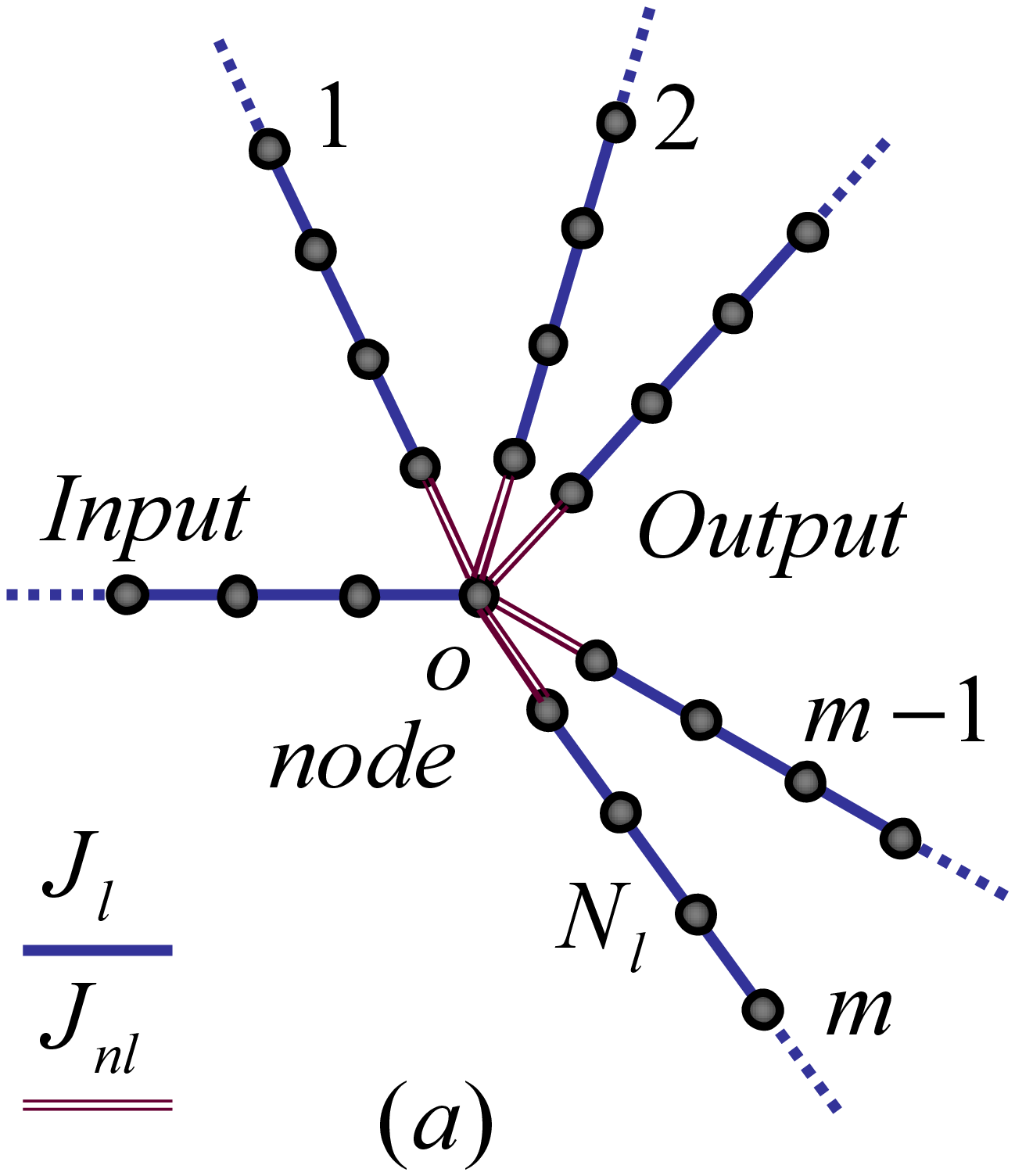} %
\includegraphics[bb=90 285 490 740, width=4 cm,clip]{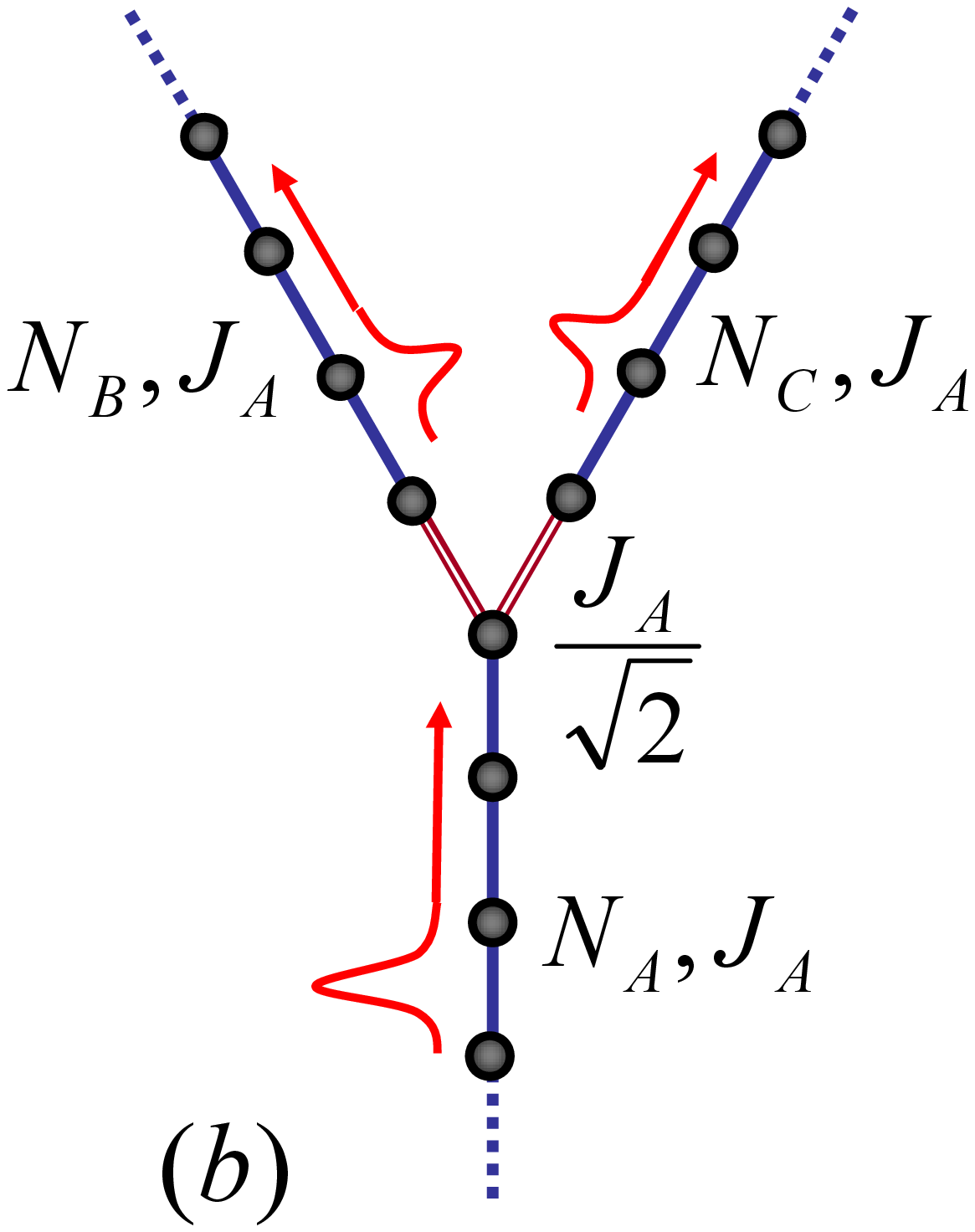}
\caption{\textit{(Color on line) (a) The star-shaped spin network with an
input spin chain $A$ and $m$ output spin chain. (b) $Y$-shaped network or
called $Y$-beam, a special star-shaped spin network, serves as the
fundamental block for the architecture of complicated quantum spin networks }
}
\end{figure}
%%%%%%%%%%%%%%%%%%%%%%%%%%%%%%%%%%%%%%%%%%%%%%%%%%%%%%%%%%%%%%%%%%%%

The basic element of an arbitrary spin network is the coupling between
spins, which is usually described by the $XY$ Hamiltonianin%
\begin{equation}
H^{XY}=J_{ij}\sum_{i,j}(S_{i}^{+}S_{j}^{-}+H.c.),
\end{equation}%
where $S_{i}^{\pm }$ are the Pauli spin operators acting on the internal
space of electron on the $i$th site. One can apply magnetic field $B_{z}$
beforehand and then switch off it, to prepare a polarized initial states
with all spins down for the quantum network. The dynamics of the lower
excitations (magnons) from this polarized state is attractive because of
their relevance to quantum information applications. In the low-temperature
and low magnon density limit, the magnon can be regarded as boson by the
Holstein-Primakoff transformation $S_{j}^{+}\simeq b_{j}^{\dag }$. Then one
can translates a $XY$ spin network into the bosonic system with the
Hamiltonian $b_{a,j}^{\dag }b_{a,j+1}+h.c$, approximately. Note that such
map is exact for the case of single magnon.

\section{Star-shaped beam splitter and its reduction}

We consider a spin network of a star shape (we call the SSSN ) as shown in
Fig. 1a. Under the Holstein-Primakoff transformation the Hamiltonian of leg $%
l$ consisting of $N_{l}$ spins with $XY$ interactions can be written as%
\begin{equation}
H_{l}=H_{l}(J_{l},N_{l})\circeq J_{l}\sum_{j=1}^{N_{l}-1}(b_{l,j}^{\dag
}b_{l,j+1}+H.c.),
\end{equation}%
where $b_{l,j}^{\dag },b_{l,j}$ are the boson operators on the $j$th site of
the $l$th leg. Here we have assumed that the couplings $J_{l}$ are the same
for a given bosonic chain $l$. We uniquely denote the Hamiltonian by $%
H_{l}(J_{l},N_{l})$ for the bosonic chain hereafter. The SSSN is constructed
by linking the $m$ output bosonic chains to the one end (or the node) $O$ of
the input leg $A$ by the couplings $J_{nl}$. The Hamiltonian of an SSSN is
of the form as the same as Eq. (2) except for the part around the node $O$.

We will show that, due to the quantum interference mechanism, by some
constrain for the coupling constants $J_{l}$\ and $J_{nl}$, an SSSN can be
reduced into an imaginary linear bosonic chain with homogeneous coupling
plus a smaller complicated network if only the single-magnon case is
concerned. The fact that the input chain $A$ is a part of this virtual
linear chain implies that the bosonic wave packet can perfectly propagate in
this virtual linear chain without the reflection by the node. This indicates
that there is a coherent split of the input bosonic wave packet because the
magnon excitation in this virtual chain actually is just a superposition of
magnon excitations in the $m$ bosonic chains.

To sketch the central idea, we first consider a special SSSN, which consists
of $m$ identical \textquotedblleft output\textquotedblright\ chains $%
B_{1},B_{2},\cdots ,$ $B_{m}$ with homogeneous coupling $J_{l}=J$, $%
J_{nl}=J_{n}$ and the same length $N$, while the length of chain $A$ is $M$.
The Hamiltonian
\begin{equation}
H=\sum_{p=1}^{m}H_{B_{p}}(J,N)+H_{A}(J,M)+H_{nod}
\end{equation}%
can be explicitly written in terms of the leg Hamiltonians $H_{B_{p}}(J,N)$
and $H_{A}(J,M)$ defined by Eq. (1) and the interactions around the node $O$
\begin{equation}
H_{nod}=-J_{n}(b_{A,M}^{\dag }\sum_{p=1}^{m}b_{B_{p},1}+H.c.).
\end{equation}

Now we construct the virtual bosonic chain $a$ of length $M+N$ with the
boson operators $b_{a,j}^{\dag }=b_{A,j}^{\dag }$ $(j=1,2,\cdots ,M)$ in the
real chain $A$ and the collective operator
\begin{equation}
b_{a,M+j}^{\dag }=\frac{1}{\sqrt{m}}\sum_{p=1}^{m}b_{B_{p},j}^{\dag }
\end{equation}%
for the virtual part, where $j=1,2,$ $\cdots ,N$. There exist $m-1$
complementary linear bosonic chains with the collective operators
\begin{equation}
b_{b_{q},j}^{\dag }=(1/\sqrt{m})\sum_{p=1}^{m}\exp (-i2\pi
pq/m)b_{B_{p},j}^{\dag }
\end{equation}%
where $p=1,2,$ $\cdots ,m,$ $q=1,2,\cdots ,$ $m-1$. It can be checked that,
together with $b_{a,j}^{\dag }$, the above defined collective operators $%
b_{a,M+j}^{\dag }$ and $b_{b_{q},j}^{\dag }$, $(q=1,2,\cdots ,$ $m-1)$ and
their conjugates also satisfy the commutative relations of boson operators.

Using operators $b_{a,M+j}^{\dag }$ and $b_{b_{q},j}^{\dag }$, we divide the
total Hamiltonian into two commutative parts
\begin{equation}
H_{b}=\sum_{q=1}^{m-1}H_{b_{q}}(J,N)
\end{equation}%
and
\begin{equation}
H_{a}=H_{a}(J,M+N)+H_{vn},
\end{equation}%
where
\begin{equation}
H_{vn}=(J-\sqrt{m}J_{n})b_{a,M}^{\dag }b_{a,M+1}+H.c.
\end{equation}%
The first Hamiltonian $H_{b}$ describes $m-1$ independent virtual bosonic
chains without input from $H_{A}$ while the second one describes a linear
bosonic chain with an impurity at the $M$th site. In usual it can reflect
the bosonic wave packet from the input leg.

Only when the coupling matching conation $J_{n}=J/\sqrt{m}$ is satisfied,
the virtual bosonic chain described by $H_{a}$ is just a standard bosonic
chain since $H_{vn}=0$. In this case no reflection occurs at the node. With
this matched node coupling, an ideal beam splitter can be realized with $m$
coherent outputs since each operator $b_{a,M+j}^{\dag }$ is a linear
combination of $b_{B_{p},j}^{\dag }$. Then it can create a superposition
from the vacuum state with bosons excitation. Each component of this
superposition represents a magnon or boson excitation in a leg. The detailed
analysis will be done with the special SSSN of $m=2$ in the next section.

\section{Y-shaped beam splitter decoupling}

Actually, we need not require the two output legs to be identical. To be
convenient, we consider the asymmetric $Y$-beam consisting of three legs $A$%
, $B$ and $C$ with three hopping integrals $J_{F}$ for $F=A,B,C$ and the
node interactions $J_{nF}$ for $F=B,C$. The total Hamiltonian reads
\begin{equation}
H=\sum_{F=A,B,C}H_{F}-\sum_{F=B,C}(J_{nF}b_{A,M}^{\dag }b_{F,1}+H.c.)
\end{equation}%
where $H_{F}=H_{F}(J_{F},N_{F})$ and $N_{A}=M,$ $N_{B}=N_{C}=N.$

In order to decouple this $Y$-beam\emph{\ }as two virtual linear bosonic
chains, we need to optimize the asymmetric couplings so that the perfect
transmission can occur in the decoupled linear bosonic chains. To this end
we introduce two sets of operators by
\begin{eqnarray}
&&b_{a,M+j}^{\dag }=\cos \theta b_{B,j}^{\dag }+\sin \theta b_{C,j}^{\dag };
\notag \\
&&b_{b,j}^{\dag }=\sin \theta b_{B,j}^{\dag }-\cos \theta b_{C,j}^{\dag },
\end{eqnarray}%
for $j=1,2,$ $\cdots ,N$. A straight forward calculation shows that the two
sets of operator act as boson operators and commutative. Here, the mixing
angle $\theta $ is to be determined as follows by the optimization for
quantum information transmission. In comparison with the optical beam
splitter, the above equation can be regarded as a fundamental issue for the
boson beam splitter.
%%%%%%%%%%%%%%%%%%%%%%%%%%%%%%%%%%%%%%%%%%%%%%%%%%%%%%%%%%%%%%%%%%%%
\begin{figure}[tbp]
\includegraphics[bb=5 315 550 810, width=4 cm,clip]{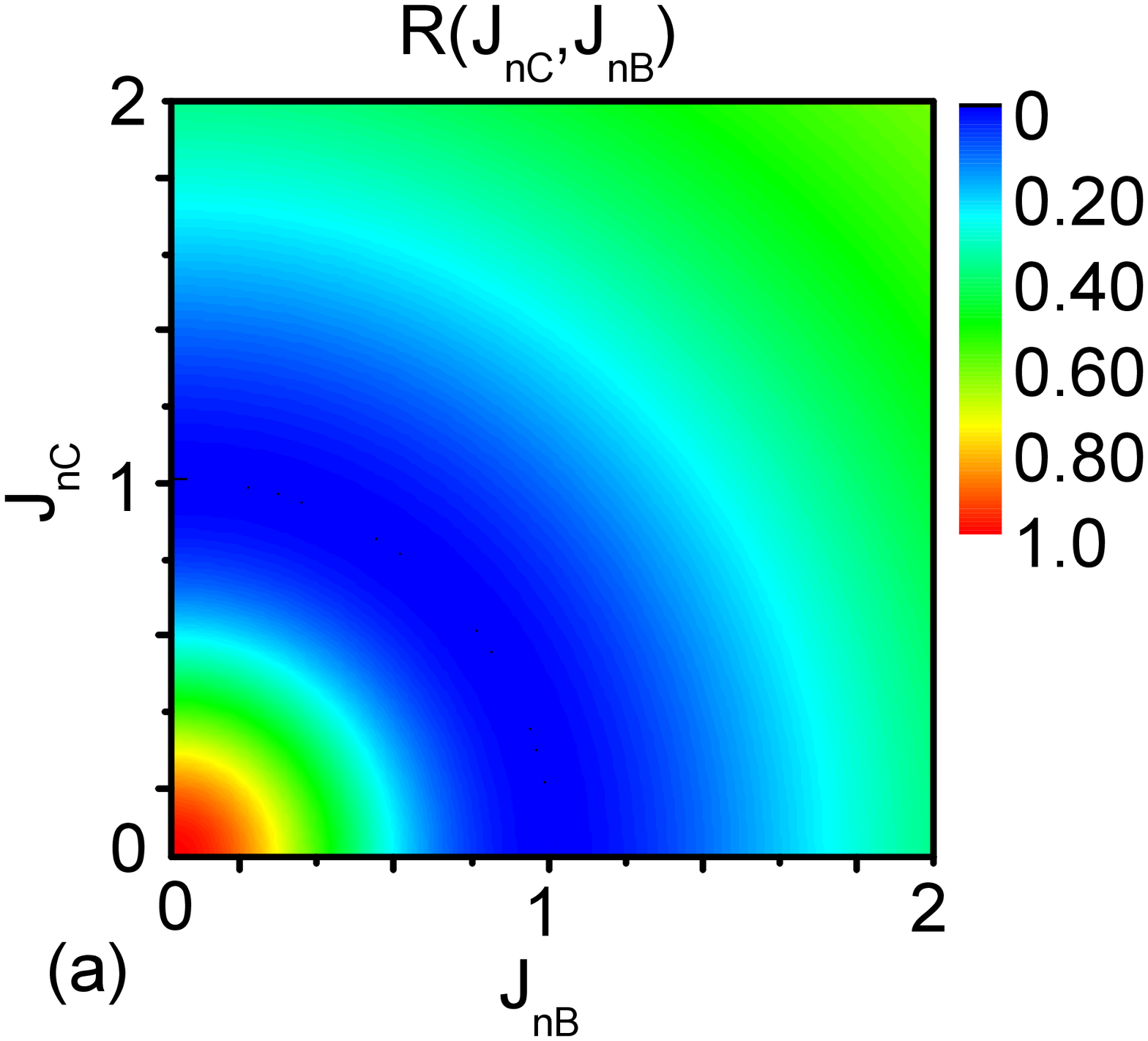} %
\includegraphics[bb=5 315 550 810, width=4 cm,clip]{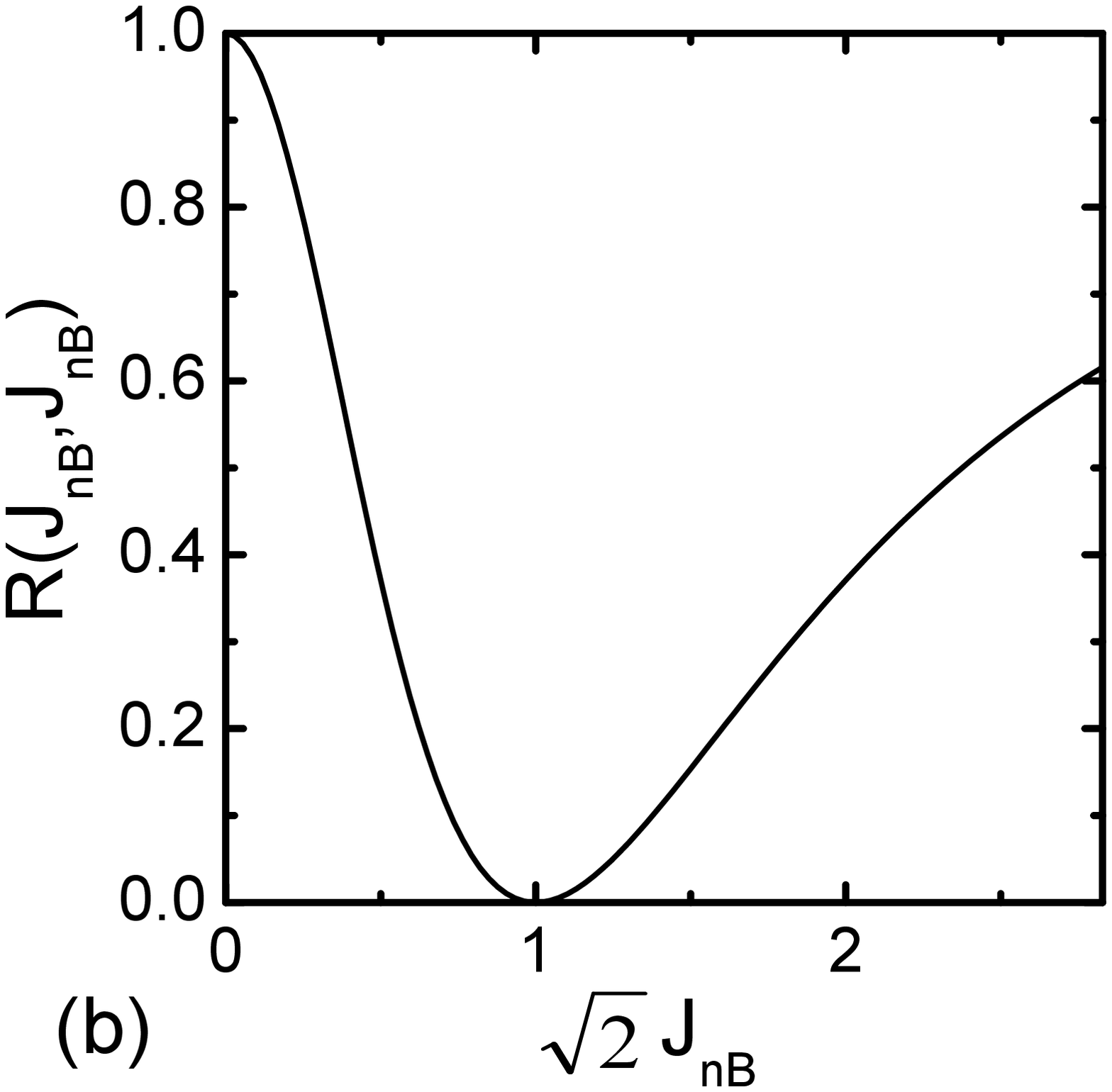}
\caption{\textit{(Color on line) (a) The contour map of the reflection
factor $R(J_{nC},J_{nB})$ as a function of $J_{nC},J_{nB}$ for the GWP with $%
\protect\alpha =0.3$ and momentum $\protect\pi /2$ in a finite system with $%
N_{A}=$$N_{B}=$$N_{C}=50$. It shows that around the matching
condition, i.e, the circle $J_{nC}^{2}+J_{nB}^{2}=J_{A}^{2}$, the
reflection factor approaches zero. (b) The profile of
$R(J_{nC},J_{nB})$ along $J_{nC}=J_{nB}$. }}
\end{figure}
%%%%%%%%%%%%%%%%%%%%%%%%%%%%%%%%%%%%%%%%%%%%%%%%%%%%%%%%%%%%%%%%%%%%

Together with the original boson operator $b_{a,j}^{\dag }=b_{A,j}^{\dag }$
for the input leg, the set with $b_{a,M+j}^{\dag }$ defines a new linear
chain $a$ with the effective couplings $J_{aj}=J_{A}$ $(j\in \lbrack 1,M-1])$%
, $J_{aM}=J_{nB}\cos \theta $ $+J_{nC}\sin \theta $ and $J_{a,M+j}=J_{B}\cos
^{2}\theta $ $+J_{C}\sin ^{2}\theta $, for $j\in \lbrack 1,$ $N-1]$. Another
virtual linear chain $b$ is defined by $b_{b,j}^{\dag }$ with homogeneous
couplings $J_{bj}=J_{B}\sin ^{2}\theta $ $+J_{C}\cos ^{2}\theta $ for $j\in
\lbrack 1,N-1]$.

In general, these two linear chains do not decouple with each other since
there exists a connection interaction around the node
\begin{eqnarray}
H_{con} &=&g(b_{b,j}^{\dag }b_{a,M+j+1}+b_{a,M+j}^{\dag }b_{b,j+1}+H.c)
\notag \\
&-&J_{AB}(b_{a,M}^{\dag }b_{b,1}+H.c.)
\end{eqnarray}%
where $g=(J_{B}-J_{C})\sin 2\theta /2$ and $J_{AB}=J_{nB}\sin \theta $ $%
-J_{nC}\cos \theta $. Fortunately, the two bosonic chains decouple with each
other when we optimize the mixing angle $\theta $ and the inter-chain
coupling by setting them as $\tan \theta =J_{nC}/J_{nB}$, $J_{B}=J_{C}$ and
then $J_{aM}=J_{nB}/\cos \theta $. Thus when we set $J_{nB}=J_{A}\cos \theta
$, the coupling matching\ condition
\begin{equation}
J_{A}=\sqrt{J_{nC}^{2}+J_{nB}^{2}}=J_{B}=J_{C}  \label{matching}
\end{equation}%
holds. Especially, the virtual bosonic chain $a$ becomes homogenous when
condition (\ref{matching})\ is satisfied. Then it can be employed to
transfer the quantum state without reflection on the node in the transformed
picture. By transforming back to the original picture, the quantum state
transfer is shown to be a perfect beam splitting. Similar to the point of
view of linear optics, such beam splitting process can generate
entanglement. We will show that the values of $J_{nB}$ and $J_{nC}$ can
determine the amplitudes of the bosonic wave packet on legs $B$ and $C$.

Now we apply the beam splitter to a special spin wave packet, a Gaussian
wave packet (GWP) with momentum $\pi /2$, which has the form
\begin{equation}
\left\vert \psi _{A\frac{\pi }{2}}(N_{0})\right\rangle =\frac{1}{\sqrt{%
\Omega _{1}}}\sum_{j}e^{-\frac{^{\alpha ^{2}}}{2}(j-N_{0})^{2}}e^{i\frac{\pi
}{2}j}\left\vert j\right\rangle   \label{GWP}
\end{equation}%
at $t=0$, where $\Omega _{1}$ is the normalization factor and $N_{0}$ is the
initial central position of the GWP at the input chain $A$. The single
excitation basis vector $\left\vert j\right\rangle =S_{A,j}^{+}|d\rangle $
is defined by the polarized state $|d\rangle $ with all spins aligned down.
As mentioned in the introduction, the conclusion we obtained for bosonic
system is exact for the single-magnon case.\ It is known from the previous
work \cite{YS1} that such GWP can approximately propagate along a homogenous
chain without spreading. Then at a certain time $t$, such GWP evolves into
\begin{equation}
\left\vert \Phi (t)\right\rangle =\cos \theta \left\vert \psi _{B\frac{\pi }{%
2}}(N_{t})\right\rangle +\sin \theta \left\vert \psi _{C\frac{\pi }{2}%
}(N_{t})\right\rangle
\end{equation}%
where $N_{t}=N_{0}+2tJ_{A}-M$, i.e., the beam splitter can split the GWP
into two cloned GWPs completely.

In order to verify the above analysis, a numerical simulation is performed
for a GWP with $\alpha =0.3$ in a finite system with $N_{A}=N_{B}=$$N_{C}=50$%
. Let $\left\vert \Phi (0)\right\rangle $ be a normalized initial state.
Then the reflection factor at time $t$ can be defined as%
\begin{equation}
R(J_{nC},J_{nB},t)=\sum_{j\in D^{\prime }}\left\vert \left\langle
j\right\vert e^{-iHt}\left\vert \Phi (0)\right\rangle \right\vert ^{2}
\end{equation}%
to depict the reflection at the node where $D^{\prime =}[1,M-1].$ At an
appropriate instant $t_{0}$, $R(J_{nC},J_{nB})=$ $R(J_{nC},J_{nB},t_{0})$ as
a function of $J_{nC}$ and $J_{nB}$ is plotted in Fig. 2. It shows that
around the coupling matching condition (\ref{matching}), the reflection
factor vanishes, which is just in agreement with our analytical result.
%%%%%%%%%%%%%%%%%%%%%%%%%%%%%%%%%%%%%%%%%%%%%%%%%%%%%%%%%%%%%%%%%%%%
\begin{figure}[tbp]
\includegraphics[bb=5 315 550 810, width=4 cm,clip]{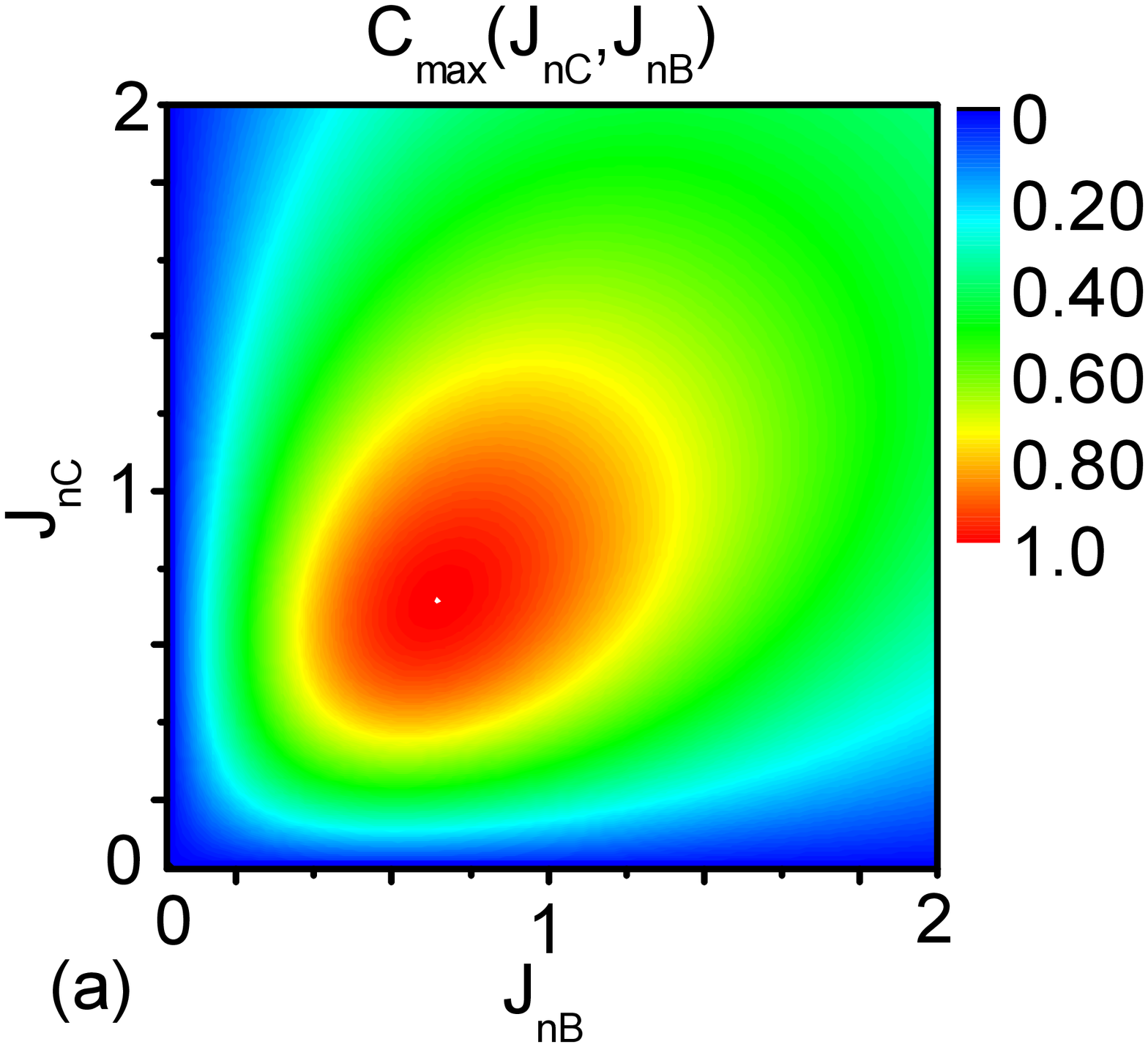} %
\includegraphics[bb=5 315 550 810, width=4 cm,clip]{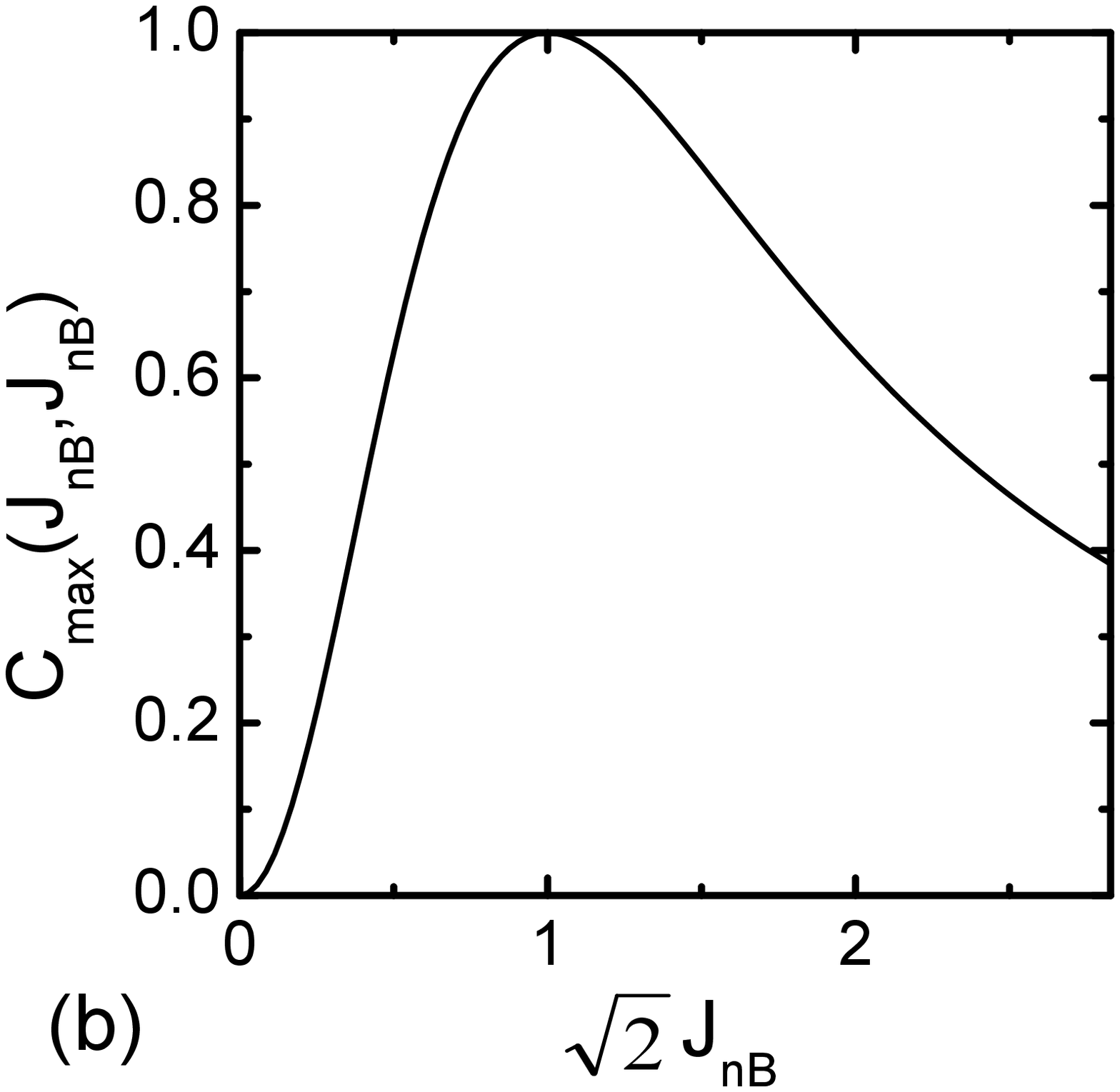}
\caption{\textit{(Color on line) (a) The contour map of maximal concurrence
of two GWPs at two legs $A$ and $B,$ $C_{\max }(J_{nC},J_{nB})$ for the same
setup as that in Fig. 2. It is found that two GWPs yield the maximal
entanglement at the point $J_{nC}=$$J_{nB}=$$J_{A}/\protect\sqrt{2}$. (b)
The profile of $C_{\max }(J_{nC},J_{nB})$ along $J_{nC}=J_{nB}$.}}
\end{figure}
%%%%%%%%%%%%%%%%%%%%%%%%%%%%%%%%%%%%%%%%%%%%%%%%%%%%%%%%%%%%%%%%%%%%

\section{Dynamic of beam splitter as entangler}

Now we consider how the SSSN can behave as an entangler to produce spin
entanglement with the $Y$-beam\textbf{\ }as an illustration. Let the input
state $|\phi (0)\rangle $ to be a single magnon excitation state in the leg $%
A$ (e.g., $=S_{A,j}^{+}|d\rangle $ or $\left\vert \psi _{A\frac{\pi }{2}%
}(N_{0})\right\rangle $ introduced by E.q. (\ref{GWP})). It can
propagate into the legs $B$ and $C$ through the node with some
reflection. On the other hand, the spin wave can be regarded as
being transferred along the virtual legs $a$ and $b$. Once we
manipulate the coupling constants to satisfy the coupling matching
condition, the spin wave can only enter the leg $a$ rather than $b$
without any reflection. Then the final state is of the magnon
excitation only in the leg $a$. As an illustration the single-magnon
excitation can be described by the state $|\phi (t)\rangle
=S_{a,M+j}^{+}|d\rangle $ or
\begin{equation}
\langle d_{A}|\phi (t)\rangle =\cos \theta |u_{jB}\rangle \otimes
|d_{C}\rangle +\sin \theta |d_{B}\rangle \otimes |u_{jC}\rangle
\end{equation}%
where $|u_{jF}\rangle =$ $S_{F,j}^{+}|d_{F}\rangle $ ($F=B,C$) represents
one-magnon excitation from the polarized state $|d_{F}\rangle $ of the chain
$F$ with all spin down. This is just an entangled state and then the $Y$%
-beam acts as an entangler similar to that in quantum optical systems.

To quantitatively characterize entanglement of two separated waves $%
\left\vert \psi _{B,C\frac{\pi }{2}}(N_{t})\right\rangle $ obtained by the
beam splitter, the total concurrence with respect to the two wave packets
located at the ends of legs $B$ and $C$ can be calculated as
\begin{equation}
C(t)=\sum_{i\in D}\left\vert \left\langle \Phi (t)\right\vert
(S_{B,i}^{+}S_{C,i}^{-}+S_{B,i}^{-}S_{C,i}^{+})\left\vert \Phi
(t)\right\rangle \right\vert
\end{equation}%
according to refs.\cite{wang,qian}. Here the domain $D=[N$ $-W,N]$ and $W=4%
\sqrt{\ln 2}/\alpha $ is the width of the wave packet. On the other hand,
the concurrence is also the function of $J_{nC}$ and $J_{nB}$. Numerical
simulation is performed for a GWP with $\alpha =0.3$ and momentum $\pi /2$
in a finite system with $N_{A}=50,$ $N_{B}=50,$ and $N_{C}=50$. The maximal
concurrence $C_{\max }(J_{nC},J_{nB})=$ $\max \{C(t)\}$ as a function of $%
J_{nC}$ and $J_{nB}$ is plotted in Fig. 3. It shows that two split wave
packets yield the maximal entanglement just at the coupling matching point $%
J_{nC}=J_{nB}$ $=J_{A}/\sqrt{2}$.

\section{Quantum interferometer for spin wave}

Finally, we consider in details a more complicated SSSN than the $Y$-beam,
the quantum interferometer for spin wave, which consists of two $Y$-beams
(see Fig. 4a). Similar to the optical interferometer, where the polarization
of photon is utilized to encode information, the SSSN uses the spin down and
up to encode the quantum information.
%%%%%%%%%%%%%%%%%%%%%%%%%%%%%%%%%%%%%%%%%%%%%%%%%%%%%%%%%%%%%%%%%%%%
\begin{figure}[tbp]
\includegraphics[bb=35 345 560 600, width=6 cm,clip]{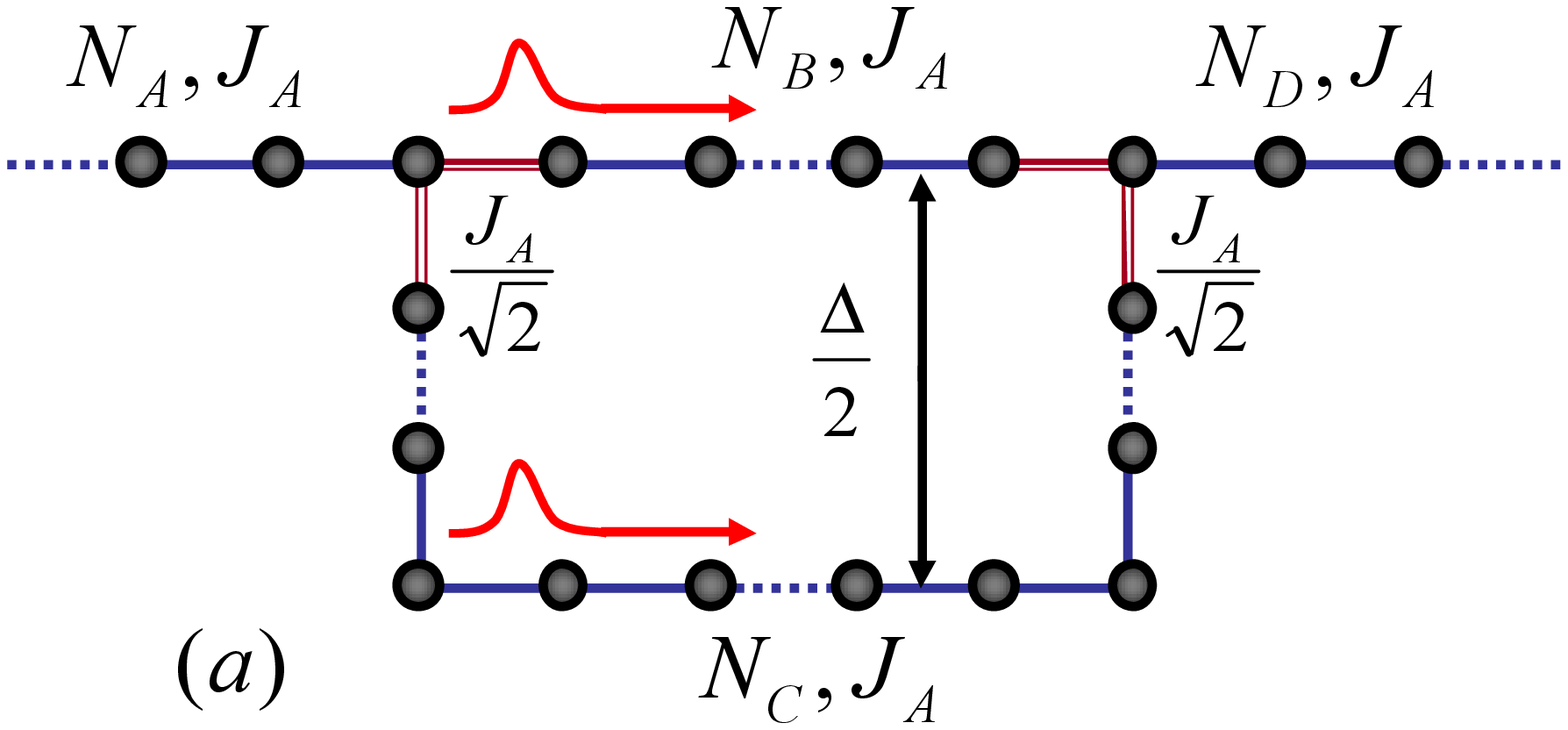} %
\includegraphics[bb=40 445 530 690, width=6 cm,clip]{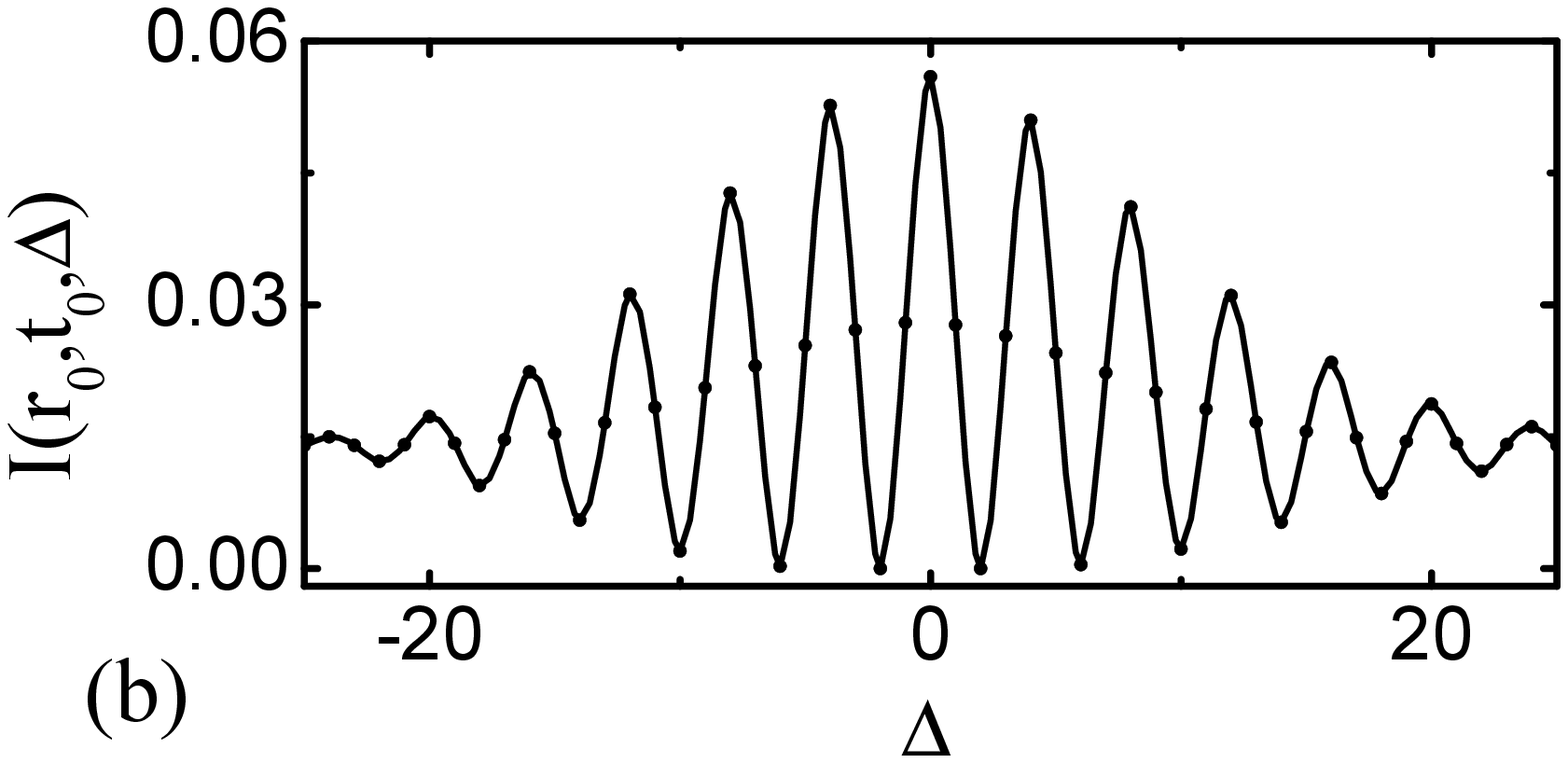}
\caption{\textit{(Color on line) (a) The interferometric network with an
input chain $A$ and output chain $D$, which consists of two $Y$-beams. $%
\Delta $ is the \textquotedblleft optical path difference\textquotedblright\
which determines the interference pattern of output spin wave. (b) The
interference pattern of output wave in the leg $D$ ($r_{0}=50$, $%
t_{0}=100/J_{A}$) for the GWP with $\protect\alpha =0.3$ in the
interferometric network with $N_{A}=N_{B}$ $=N_{D}=50$, $N_{C}=N_{B}+\Delta $%
. }}
\end{figure}
%%%%%%%%%%%%%%%%%%%%%%%%%%%%%%%%%%%%%%%%%%%%%%%%%%%%%%%%%%%%%%%%%%%%

We still use the evolution of GWP to demonstrate the physical mechanism of
such setup. Firstly, we consider the simplest case with the path difference
(defined in Fig. 4a) $\Delta =0$. It is easy to show that such network is
equivalent to two independent virtual chains with lengths $N_{A}+N_{B}+N_{D}$
and $N_{B}$ respectively when the coupling matching condition is satisfied.
Then the initial GWP will be transmitted into the leg $D$ without any
reflection. This fact can be understood according to the interference of two
split GWPs. It means that the nonzero $\Delta $ should affect the shape of
the pattern of output wave.

Actually, from the above analysis about the GWP propagating in the $Y$-beam,
we note that the conclusion can be extended to the $Y$-beam consisting of
two different length legs $N_{B}\neq N_{C}$. It is due to the locality of
the GWP and the fact that the speed of the GWP only depends on the coupling
constant. Thus the interference pattern at site $r_{0}$ and time $t_{0}$ in
leg $D$ can be presented as $I(r_{0},t_{0},\Delta )=$ $\left\vert
\left\langle r_{0}\right\vert \exp (-iHt_{0})\left\vert \Phi
(0)\right\rangle \right\vert ^{2}$. Numerical simulation of $%
I(r_{0},t_{0},\Delta )$ for the input GWP in the interferometric network
with $N_{A}=N_{B}$ $=N_{D}=50$, $N_{C}=N_{B}+\Delta $\ is performed. For $%
r_{0}=50$, $t_{0}=100/J_{A}$, a perfect interference phenomenon by $%
I(r_{0},t_{0},\Delta )$ is observed for the range $\Delta \in \lbrack
-25,25] $ in Fig. 4b.

In summary, we point out that our coherent quantum network for spin wave can
be implemented by an array of quantum dots or other artificial atoms. It
will enable an elementary quantum device for scalable quantum computation,
which can coherently transfer quantum information among the qubits to be
integrated. The observable effects for spin wave interference may be
discovered in the dynamics of magnetic domain in some artificial quantum
material.

This work is supported by the NSFC with grant Nos. 90203018, 10474104 and
60433050; and by the National Fundamental Research Program of China with
Nos. 2001CB309310 and 2005CB724508.


\begin{thebibliography}{a}
\bibitem[a]{email} emails: songtc@nankai.edu.cn \newline
and suncp@itp.ac.cn

\bibitem[b]{www} Internet www site: http://www.itp.ac.cn/\symbol{126}suncp

\bibitem{QOP} R. Loudon, \textit{The quantum theory of light,} (Oxford,
2000); M.O. Scully and M.S. Zubairy, Quantum Optics, (Oxford, 1997).

\bibitem{S-photon} J.D. Franson; Phys. Rev. A \textbf{56}, 1800-1805 (1997);
K. Jacobs and P.L. Knight; Phys. Rev. A \textbf{54}, R3738(1996); T. Wang,
M. Kostrun, and S.F. Yelin; Phys. Rev. A \textbf{70}, 053822 (2004).

\bibitem{Entng} M. Zukowski, A. Zeilinger, and M.A. Horne, Phys. Rev. A
\textbf{55}, 2564(1997); J.L. van Velsen, Phys. Rev. A \textbf{72}, 012334
(2005).

\bibitem{N-interfer} H. Rauch, W. Treimer, and U. Bonse, Phys. Lett. A
\textbf{47}, 369 (1974).

\bibitem{C-atom} D. Cassettari, B. Hessmo, R. Folman, T. Maier, and J.
Schmiedmayer, Phys. Rev. Lett. \textbf{85}, 5483(2000); U. V. Poulsen and K.
Momer, Phys. Rev. A \textbf{65}, 033613 (2002); D. C. E. Bortolotti and J.L.
Bohn;Phys. Rev. A \textbf{69}, 033607 (2004).

\bibitem{BEC} F. Burgbacher and J. Audretsch; Phys. Rev. A \textbf{60},
R3385(1999); N.M. Bogoliubov, A. G. Izergin, N.A. Kitanine, A.G. Pronko, and
J. Timonen; Phys. Rev. Lett. \textbf{86}, 4439 (2001)

\bibitem{our-spin} Y. Li, T. Shi, B. Chen, Z. Song, C.P. Sun, Phys. Rev. A
\textbf{71}, 022301 (2005); Z. Song, C. P. Sun, Low Temperature Physics
\textbf{31}, 686 (2005).

\bibitem{our-bloch} T. Shi, Y. Li, Z. Song, and C.P. Sun, Phys. Rev. A
\textbf{71}, 032309 (2005); Y.Li, Z.Song, and C.P. Sun, e-print
quant-ph/0504175.

\bibitem{YS1} S. Yang, Z. Song, and C.P. Sun, Phys. Rev. A \textbf{73},
022317 (2006).

\bibitem{Plenio} M.B. Plenio, J. Hartley and J. Eisert, New J. Phys. \textbf{%
6}, 36 (2004); A Perales, M B. Plenio, quant-ph/0510105.

\bibitem{du} Q. Chen, J. Cheng, K-L. Wang, J. Du, quant-ph/0510147.

\bibitem{Div} D. P. DiVincenzo, Fortsch. Phys. \textbf{48}, 771 (2000) (in
special issue on Experimental Proposals for Quantum Computation).

\bibitem{Bose1} S. Bose, Phys. Rev. Lett. \textbf{91}, 207901 (2003).

\bibitem{Bose2} M-H. Yung and S. Bose, Phys. Rev. A. \textbf{71}, 032310
(2005).

\bibitem{Ekert1} M. Christandl, N. Datta, A. Ekert, and A. J. Landahl, Phys.
Rev. Lett. \textbf{92}, 187902 (2004).

\bibitem{Ekert2} C. Albanese, M. Christandl, N. Datta, and A. Ekert, Phys.
Rev. Lett. \textbf{93}, 230502 (2004).

\bibitem{wang} X. Wang and P. Zanardi, Phys. Lett. A \textbf{301}, 1 (2002);
X. Wang, Phys. Rev. A \textbf{66}, 034302 (2002).

\bibitem{qian} X-F. Qian, Y. Li, Y. Li, Z. Song, and C.P. Sun, Phys. Rev. A
\textbf{72}, 062329 (2005).
\end{thebibliography}
\end{document}